# Average Transmission Decrease in One-Dimensional Photonic Structures by Widening the Random Layer Thickness Distribution

Francesco Scotognella[1,2]*

[1] Politecnico di Milano, Dipartimento di Fisica and Istituto di Fotonica e Nanotecnologie CNR, Piazza Leonardo da Vinci 32, 20133 Milano, Italy
[2] Center for Nano Science and Technology@PoliMi, Istituto Italiano di Tecnologia, Via Giovanni Pascoli, 70/3, 20133, Milan, Italy
* francesco.scotognella@polimi.it

**Abstract**
In this work we have studied the optical properties of disordered photonic structures, in which we have controlled the distribution of the random layer thickness. Such structures are characterized by an alternation of high and low refractive index layers, but the layer thicknesses follow the aforementioned distributions. We have used two types of distribution: a distribution in which each thickness has the same probability to occur and one in which the thickness follows a Gaussian function. We have simulated the average transmission all over the spectrum for photonic structure characterized by a different width of the distribution. We have found that the choice of the distribution of the layer thickness is a control of the average transmission of a random photonic structure.

**Introduction**
Periodicity in one dimensional multilayer photonic structures is interrupted by introducing different materials, by randomizing the sequence of the layers, and by randomizing the thickness of the layers [1–3]. Another type of non-periodic photonic structures are quasicrystals [4–6] and, the breaking of the periodicity in photonic structures is widely studied also in two and three dimensions [7–9]. While periodic photonic crystals show a characteristic photonic band gap [10,11], with a spectral position that follows the Bragg-Snell law [12,13], the disordered structures show randomly arranged transmission depths in the transmission spectrum [1,14–16]. The disordered photonic structures can be employed for the realization of random lasers [2,17–19] and for light trapping in photovoltaic cells [20].
It has been theoretically proposed [21] and experimentally demonstrated [22] that the average transmission of a photonic structure characterized by a random layer thickness is lower with respect to the one of a periodic photonic crystal. But in these two studies an analysis of the influence of the distribution of the random layer thicknesses on the optical properties of the disordered structures is missing.
In this work we have studied, by using the transfer matrix method, the transmission spectrum of the disordered structures in which the layer thickness follows two types of distributions. The first distribution results in a range of thicknesses with the same probability to occur, while the second distribution follows a Gaussian function. The simulations take into account the Sellmeier equation of Silicon dioxide and Zirconium dioxide. We show that the choice of the distribution strongly affects the optical properties of the structures.

**Methods**
All the one-dimensional photonic structures studied in this work are composed by 28 alternating layers of Silicon dioxide and Zirconium dioxide, respectively. The dispersion formula (Sellmeier equation) of the refractive index of Silicon dioxide is

$$n_{SiO_2}^2(\lambda) - 1 = \frac{0.6961663\lambda^2}{\lambda^2 - 0.0684043^2} + \frac{0.4079426\lambda^2}{\lambda^2 - 0.1162414^2} + \frac{0.8974794\lambda^2}{\lambda^2 - 9.896161^2}$$

as reported in [23], while the formula for Zirconium dioxide is

$$n_{ZrO_2}^2(\lambda) - 1 = \frac{1.347091\lambda^2}{\lambda^2 - 0.062543^2} + \frac{2.117788\lambda^2}{\lambda^2 - 0.166739^2} + \frac{9.452943\lambda^2}{\lambda^2 - 24.320570^2}$$

as reported in [24].
The thicknesses of the layers are random, following two types of distributions. In the first case, the thickness of the $i$th layer is $d_i = d \pm s$, where $d = 100$ nm and $s$ is an integer that spans from 0 to $\eta/2$ ($\eta$ is an even number). For example, if $\eta = 20$, the random integer spans from 0 to 10. Since in this distribution all the layer thicknesses in the range $\eta$ have the same probability to occur, we will call it flat distribution. In the second case, the probability of a random integer to occur follows a Gaussian function, centred at 100 nm and with a standard deviation $\sigma$. We simulate transmission spectra for different $\eta$ and for different $\sigma$.

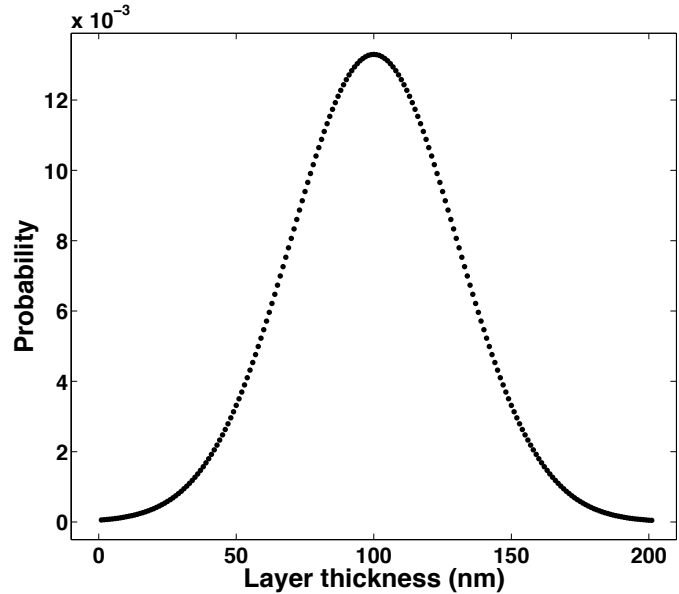

**Figure 1.** Gaussian function, with unitary area, centred at 100 nm and with $\sigma = 30$. The Gaussian function is truncated to the range [1-200].

As shown in Figure 1, a Gaussian function, with $\sigma = 30$, results in a probability for a 100 nm layer thickness of about 1.33%, while the probability to have a 90 nm thick layer is about 1.26%, for 70 nm the 0.55% etc. The Gaussian function is truncated to the range [1-200].
To simulate the transmission spectra of the photonic structures, we have employed the transfer matrix method [25–27]. We have integrated the transmission all over a range between 300 and 1200 nm (step of 1 nm), normalizing to the transmission of a perfectly transparent material (i.e. 100% transmission at any wavelength). For each $\eta$ or $\sigma$, we have simulated 1000 different thickness random sequences.

**Results and Discussion**
First, we have calculated the transmission spectra of random photonic structures with layer thicknesses following the flat distribution. For the readers information, the transmission spectrum of the periodic photonic crystal with the same parameters shows a photonic band gap centred at about 730 nm with a full width at half maximum (FWHM) of 190 nm. Instead, the disordered structure corresponding to $\eta$ = 120 shows very intense transmission depths all over the studied spectral region (Figure 2a). In Figure 2b we show the histograms of the normalized average transmission for different values of $\eta$.

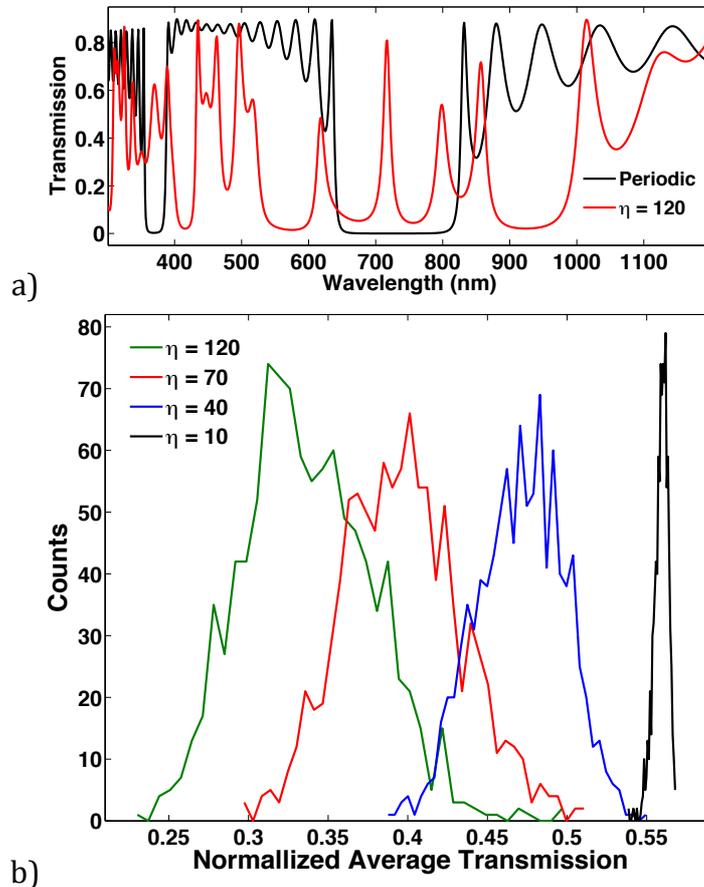

**Figure 2.** a) Transmission spectra of a periodic photonic crystal (black curve) and of a disordered photonic structure corresponding to $\eta$ = 120. b) Histograms of the normalized average transmission, in the range 300 – 1200 nm, for different values of $\eta$ of the distribution of the layer thickness.

We observed that the normalized average transmission decreases by increasing $\eta$. Concomitant with the normalized average transmission lowering, an decrease of the deviation from the mean value occurs (the histograms become narrower with increasing $\eta$). These findings are very clear looking at the trend of the normalized average transmission as a function of $\eta$.

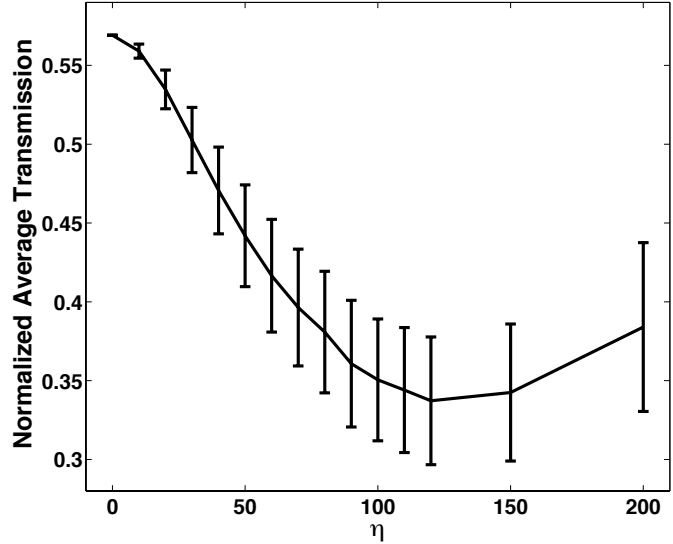

**Figure 3.** Normalized average transmission as a function of the value of $\eta$. The error bars correspond to the standard deviation of such values.

The values, together with the standard deviation of the normalized average transmission, are reported in Table 1. With $\eta = 0$ we have referred to the periodic structure, in which all the layers are 100 nm thick.

**Table 1.** Normalized average transmission, with the corresponding standard deviation, as a function of the value of $\eta$.

| $\eta$ | Normalized average transmission | Standard deviation |
|---|---|---|
| 0 (Periodic) | 0.5691 | - |
| 10 | 0.5590 | 0.0044 |
| 20 | 0.5348 | 0.0123 |
| 30 | 0.5027 | 0.0207 |
| 40 | 0.4707 | 0.0275 |
| 50 | 0.4419 | 0.0323 |
| 60 | 0.4166 | 0.0358 |
| 70 | 0.3964 | 0.0370 |
| 80 | 0.3808 | 0.0386 |
| 90 | 0.3608 | 0.0402 |
| 100 | 0.3505 | 0.0386 |
| 110 | 0.3440 | 0.0396 |
| 120 | 0.3372 | 0.0405 |
| 150 | 0.3424 | 0.0435 |
| 200 | 0.3840 | 0.0535 |

The decrease of the normalized average transmission is quite remarkable for large values of $\eta$. For example, the normalized average transmission for $\eta = 200$ is the 59% of the one of the periodic structure, testifying the significant role played by the non uniformity of the layer thickness. The increase of such non uniformity is also concomitant to an increase of the standard deviation of the normalized average transmission.

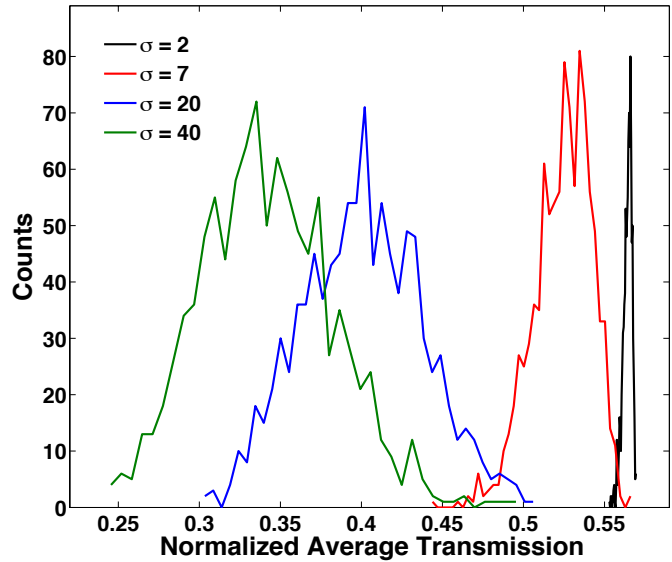

**Figure 4.** Histograms of the normalized average transmission, in the range 300 – 1200 nm, for different values of $\sigma$ of the Gaussian distribution of the layer thickness.

With the Gaussian distribution of the layer thickness, the trend is very similar to the one of the flat distribution. In Figure 4 we show the histograms of the normalized average transmission for some different values of $\sigma$, and the trend in Figure 5 (with the values shown in Table 2). For both the distributions we observe a trend that shows a minimum (for $\eta = 120$ and for $\sigma = 40$).

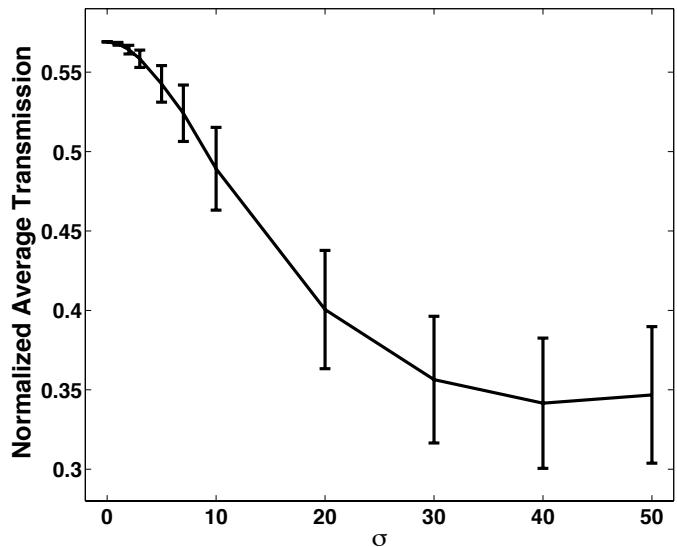

**Figure 5.** Normalized average transmission as a function of the value of $\sigma$. The error bars correspond to the standard deviation of such values.

**Table 2.** Normalized average transmission, with the corresponding standard deviation, as a function of the value of $\sigma$.

| $\sigma$ | Normalized average transmission | Standard deviation |
|---|---|---|
| 0 (Periodic) | 0.5691 | - |
| 1 | 0.5679 | 0.0009 |
| 2 | 0.5642 | 0.0027 |
| 3 | 0.5584 | 0.0055 |
| 5 | 0.5426 | 0.0115 |
| 7 | 0.5242 | 0.0178 |
| 10 | 0.4892 | 0.0261 |
| 20 | 0.4005 | 0.0372 |
| 30 | 0.3564 | 0.0399 |
| 40 | 0.3416 | 0.0410 |
| 50 | 0.3468 | 0.0430 |

A simulation with a Gaussian distribution of the layer thickness has been already reported in Ref. [21], in which a non negligible reflectance (i.e., a transmission decrease) at all the wavelengths of the spectrum has been observed. We here show that the choice of the distribution width, quantified by $\sigma$, results in a control of this transmission lowering in a spectral range.

**Conclusion**
In this work we have studied one dimensional photonic structures, characterized by a periodic alternation of Silicon dioxide and Zirconium dioxide, but also by a random distribution of the layer thickness. We have used a flat distribution, in which each thickness has the same probability of occur, and a distribution in which the probability of a certain thickness to occur follows a Gaussian function. We have simulated the transmission spectra with the transmission matrix method, and we have calculated the average transmission of photonic structures characterized by a different width of the distributions. We show that a control on the average transmission of a random photonic structure is the choice of the distribution of the layer thickness.

**Acknowledgement**
The author thanks Ilka Kriegel for revising the manuscript.